\documentclass[aps,preprint,prd,showpacs,nofootinbib]{revtex4-1}
\usepackage{graphicx}
\usepackage{bm}
\usepackage{times}
\usepackage{hyperref}
\usepackage{slashed}
\usepackage{amsmath}
\usepackage{xcolor}
\usepackage{fancyhdr}
\usepackage{footmisc}
\usepackage{soul}
\usepackage{lipsum}
\usepackage{multirow}
\pdfoutput=1
\usepackage{amssymb}
\usepackage{hyperref}
\newcommand{\be}{\begin{equation}}
\newcommand{\ee}{\end{equation}}
\newcommand{\bea}{\begin{eqnarray}}
\newcommand{\eea}{\end{eqnarray}}
\newcommand{\bes}{\begin{subequations}}
\newcommand{\ees}{\end{subequations}}

\newcommand{\bc}{\begin{center}}
\newcommand{\ec}{\end{center}}
\usepackage{amsmath, amssymb} 
\usepackage{slashed}
\usepackage{natbib}

\begin{document}
\title{ Introducing scalar leptoquarks into a 3-3-1 model to solve the $(g-2)_\mu $ puzzle}
\author{A. Doff$^b$}\email{agomes@utfpr.edu.br}  
\author{C. A. de S. Pires$^{a}$}\email{cpires@fisica.ufpb.br} 
\affiliation{$^{a}$Departamento de F\'isica, Universidade Federal da Para\'iba, Caixa Postal 5008, 58051-970, Jo\~ao Pessoa, PB, Brazil} 
\affiliation{$^{b}$Universidade Tecnologica Federal do Parana - UTFPR - DAFIS, R. Doutor Washington Subtil Chueire, 330 - Jardim Carvalho,  84017-220, Ponta Grossa, PR, Brazil}

\date{\today}
\vspace{1cm}
\begin{abstract}
In this work we introduce scalar leptoquarks into the 3-3-1 model with right-handed neutrinos with the aim of solving the  $(g-2)_{\mu}$ puzzle. We show that besides the model supports leptoquarks in the  octet, sextet, triplet and singlet representations, we identified that  only one specif  leptoquark in the singlet representation  leads to flip of chirality as required to generate positive and robust contribution to the $(g-2)_\mu$. Then we calculate its contributions to  $(g-2)_\mu$ and  to the decay process  $\mu \rightarrow e \gamma$ and discuss the results.
\end{abstract}

\maketitle
\section{Introduction}

\par The anomalous magnetic moment of the muon, $ a_\mu =(\frac{g-2}{2})_\mu $ , has long stood as a tantalizing puzzle in the  particle physics\cite{Muong-2:2001kxu,Jegerlehner:2017gek}. Experimental measurements of $ a_\mu$ have consistently shown a  deviation from the predictions of the Standard Model (SM) of the order of the electroweak contribution\cite{Muong-2:2002wip},
\begin{eqnarray}
  &&  a_\mu^{Exp}=116592059(22)\times 10^{-11},\nonumber \\
  &&a_\mu^{SM}=  116591802(2) \times 10^{-11},\nonumber \\
  &&\Delta a_\mu=a_{\mu}^{Exp}-a_\mu^{SM}=251 ( 59) \times 10^{-11}.
\end{eqnarray}
Such result has sparked an  intense theoretical and experimental scrutiny to unravel its underlying causes\cite{Lindner:2016bgg,Aoyama:2020ynm,Athron:2021iuf}.

\par One intriguing avenue of investigation  of $a_\mu$ lies in the realm of leptoquark models. Recent theoretical studies have demonstrated the viability of chirality flip effects mediated by leptoquarks(LQ) in addressing the muon anomalous magnetic moment problem, potentially reconciling the observed anomaly with experimental measurements\cite{Athron:2021iuf,Crivellin:2021rbq,Stockinger:2022ata}.

Leptoquarks are hypothetical particles in the form of  scalars or vectors   that  have gained significant attention in the last decades\cite{Dorsner:2016wpm}. These particles  combine features of both leptons and quarks and  have  fractional electric charge\cite{Pati:1974yy,PhysRevLett.32.438}. All these features have important impact in flavor physics\cite{Dorsner:2016wpm}. 

Besides the standard model of particle physics does not predict leptoquarks, they emerge as essential components in various extensions to the Standard Model, including grand unified theories (GUTs)\cite{Pati:1973uk}, supersymmetry\cite{Hewett:1988xc}, and composite models \cite{Schrempp:1984nj}. However the symmetries of the SM support the introduction of leptoquarks into it in the triplet, doublet and singlet representations. The last two representations are capables of generating flip of chirality among the fermions which generate an enhancement of $(g-2)_\mu$ by the factor  $\frac{m_t}{m\mu}$\cite{ColuccioLeskow:2016dox}. 

\par As in the SM case, the $SU(3)_C \times SU(3)_L \times U(1)_N$ (3-3-1) models face, too, serious difficulities in  accommodating the current value of $a_\mu$ at one-loop\cite{Kelso:2014qka,DeJesus:2020yqx,CarcamoHernandez:2020pxw,Ky:2000ku,deJesus:2020ngn,Pinheiro:2021mps,Hong:2022xjg}\footnote{It was showed in Ref. \cite{Pinheiro:2021mps} that the extension of the model to generate the inverse seesaw mechanism is not able to explain $\Delta a_\mu$ anomaly }. In view of this, we introduce leptoquarks to the original particle content of a version of the 3-3-1 models that features right-handed neutrinos as third component of the lepton triplets (331RHN for short)\cite{Foot:1994ym,Montero:1992jk}. As it is the first work in which leptoquarks are introduced into this model.\footnote{In fact leptoquarks were first introduced  in  \cite{Doff:2023bgy}  where it was considered the possibility of including leptoquarks in the minimal 3-3-1 model with the aim of evading the Landau pole}, we first  identify the representation content of leptoquarks the model supports. Next we  determine which leptoquark representation leads to flip of chirality among fermions in the 331RHN. We show that the model allows leptoquarks in the octet, sextet, triplet and singlet representation but only one specific leptoquark in the singlet representation is capable of generating  flip of chirality among the fermions. We then focus on this leptoquark and calculate and discuss its contributions at one loop to $(g-2)_\mu$  with the aim of obtaining which form of enhancement this leptoquark promote. We also discuss the contribution of such leptoquark to the $\mu \rightarrow e \gamma$ decay and if it  discriminata family.

This paper is organized in the following way: in Sec. II we present the main aspects of the model. In Sec. III we introduce leptoquarks identifying representation content and Yukawa interactions. In Sec. IV we review the one loop contributions of the original 331RHN to $(g-2)_\mu$ and calculate the contributions due to the singlet of leptoquark. In Sec. V we calculate the contribution of the singlet of leptoquark to the $\mu \rightarrow e \gamma$. In Sec. VI we discuss all the findings and in Sec. VII we present our conclusions.

\section{The main aspects of the model} \label{sec:2}

In the 331RHN right-handed neutrinos compose the third component of the leptonic triplets\cite{Foot:1994ym,Montero:1992jk},
\begin{equation}
L_{a_L}= \begin{pmatrix}
\nu_{a}     \\
e_{a}       \\
\nu^{c}_{a} \\
\end{pmatrix}_{L} \sim (1,3,-1/3), \quad e_{a_R}\sim (1,1,-1),
\end{equation}
with $a=1\,,\,2\,\,3$ representing the three SM generations of leptons.

In the hadronic sector anomaly cancellation demands that one family transforms differently from the other two. Here we chose the first two families of quarks  transforming as anti-triplet while the third one  transforming as triplet by $\text{SU}(3)_\text{L}$,
\begin{eqnarray}
&&Q_{i_L} = \left (
\begin{array}{c}
d_{i} \\
-u_{i} \\
d^{\prime}_{i}
\end{array}
\right )_L\sim(3\,,\,\bar{3}\,,\,0)\,,u_{iR}\,\sim(3,1,2/3),\,\,\,\nonumber \\
&&\,\,d_{iR}\,\sim(3,1,-1/3)\,,\,\,\,\, d^{\prime}_{iR}\,\sim(3,1,-1/3),\nonumber \\
&&Q_{3L} = \left (
\begin{array}{c}
u_{3} \\
d_{3} \\
u^{\prime}_{3}
\end{array}
\right )_L\sim(3\,,\,3\,,\,1/3),u_{3R}\,\sim(3,1,2/3),\nonumber \\
&&\,\,d_{3R}\,\sim(3,1,-1/3)\,,\,u^{\prime}_{3R}\,\sim(3,1,2/3),
\label{quarks} 
\end{eqnarray}
where  the index $i=1,2$ is restricted to only two generations. The negative signal in the anti-triplet $Q_{i_L}$ is just to standardise the signals of the charged current interactions with the gauge bosons.  The primed quarks are new heavy quarks with the usual $(+\frac{2}{3}, -\frac{1}{3})$ electric charges.

The original scalar sector of the  model is composed by three triplets of scalars
\begin{eqnarray}
\eta = \left (
\begin{array}{c}
\eta^0 \\
\eta^- \\
\eta^{\prime 0}
\end{array}
\right ),\,\rho = \left (
\begin{array}{c}
\rho^+ \\
\rho^0 \\
\rho^{\prime +}
\end{array}
\right ),\,
\chi = \left (
\begin{array}{c}
\chi^0 \\
\chi^{-} \\
\chi^{\prime 0}
\end{array}
\right ),
\label{scalarcont} 
\end{eqnarray}
with $\eta$ and $\chi$ transforming as $(1\,,\,3\,,\,-1/3)$
and $\rho$ as $(1\,,\,3\,,\,2/3)$. After spontaneous breaking of the 3-3-1  symmetry ($\langle \eta^0 \rangle =v_\eta\,\,,\,\, \langle \rho^0 \rangle =v_\rho\,\,,\,\,\langle \chi^{\prime 0} \rangle =v_{\chi^{\prime}}$ with $v_{\chi^{\prime}} >> v_\eta\,,\,v_\rho$ )  the scalar sector\footnote{Here we follow Ref.\cite{DeJesus:2020yqx},
 and assume $v_\eta=v_\rho=v$} get composed by three CP-even scalars, $h_1$, $h_2$, and $H$ where the first recovers the features of the standard Higgs and the other two are heavy scalars with mass proportional to $v_{\chi^{\prime}}$ with $m^2_{h_1}\approx \frac{v v_{\chi^{\prime}}}{2}$ and $m^2_H \sim v^2_{\chi^{\prime}}$. The model also has in its spectrum of scalars one pseudoscalar we call $A$, whose mass is given by $ m^2_A \approx \frac{v v_{\chi^{\prime}}}{2}$ and two singly charged scalars $h^+_1$ and $h^+_2$ with $m^2_{h^+_1} \approx v v_{\chi^{\prime}}$ and $m^2_{h^+_2} \sim v^2_{\chi^{\prime}}$. Except by $h_1$ that must recover the 125 GeV standard Higgs, all the other scalars have their masses determined by $v_{\chi^{\prime}}$ which means that the natural energy scale of the masses of these scalars is the TeV scale\footnote{For a detailed discussion of the scalar sector, see Refs. \cite{Long:1997vbr,Diaz:2003dk,Pinheiro:2022bcs}}.

Here the most economical Yukawa interactions that generate the correct mass for all standard quarks and leptons are composed by the terms
\begin{equation}\label{yukawa}
-{\cal L}_Y \supset g_{ia} \bar Q_{i_L} \eta^* d_{a_R} + h_{3a} \bar Q_{3_L} \eta u_{a_R} + g_{3a} \bar Q_{3_L} \rho d_{a_R} + h_{ia} \bar Q_{i_L} \rho^* u_{a_R} + Y\bar L \rho e_R  +\mbox{H.c.}\,,
\end{equation}
where $a=1,2,3$ and the parameters $g_{ab}$, $h_{ab}$ and $Y$ are Yukawa couplings that, for sake of simplification,  we consider reals\footnote{For a more comprehensive discussion of the Yukawa sector, see Ref. \cite{Doff:2006rt}}.

The gauge sector of the model involves nine gauge bosons related to the electroweak sector, where four of them are the standard gauge bosons ($W^{\pm}$, $ Z^0$,  $\gamma$) and the other  five   ($W^{\prime \pm}$,  $U^0$, $U^{0 \dagger}$,  $Z^{\prime}$) are the typical 3-3-1 gauge bosons, and eight gluons $g$ relaterd to the strong force. The standard gauge bosons recover the physics of the SM while the new gauge bosons have mass belonging to the TeV scale and interactions that connect the standard particle content with the specific 3-3-1 particle content\footnote{Concerning gauge bosons of the model see Ref. \cite{Long:1996rfd,Cao:2016uur}}. The mass of $W^{\prime}$ and $Z^{\prime}$ are given by,
\begin{equation}
M_{Z^{\prime}}^2 \approx \frac{g^2 c^2_W }{(3-4s_w^2)}v_{\chi^{\prime}}^2 \,\,,\,\,\,\,M_{W^{\prime}}^2 = \frac{g^2}{4}\left( \frac{v^2}{2} + v_{\chi^{\prime}}^2 \right).
\end{equation}

Observe that $v_{\chi^{\prime}}$ practically determine the masses of the spectrum of new particle characteristic of the 3-3-1.

\section{Leptoquark: representation content and Yukawa interactions}
With all this in hand  we are ready to determine the possible set of leptoquarks that the 331RHN supports. Observe that from the quark and lepton content of the model we can build the following  bilinears,
\begin{eqnarray}
    && \bar L^C_{a_L} Q_{3_L} \sim (1\,,\,3\,,\,-1/3)\times (3\,,\,3\,,\,1/3) \sim (3\,,\,3^* \oplus 6\,,\,0),\nonumber \\
    &&\bar L^C_{a_L} Q_{i_L} \sim (1\,,\,3,-1/3)\times (3\,,\,3^*\,,\,0) \sim (3\,,\,1 \oplus 8\,,\,-1/3),\nonumber \\
    &&\bar L_{a_L} d_{b_R} \sim (1\,,\,3^*,1/3)\times (3\,,\,1\,,\,-1/3) \sim (3\,,\,3^*\,,\,0),\nonumber \\
    &&\bar L_{a_L} u_{b_R} \sim (1\,,\,3^*,1/3)\times (3\,,\,1\,,\,2/3) \sim (3\,,\,3^*\,,\,1),\nonumber \\
    &&\bar e^C_{a_R} d_{b_R} \sim (1\,,\,1,-1)\times (3\,,\,1\,,\,-1/3) \sim (3\,,\,1\,,\,-4/3),\nonumber \\
    &&\bar e^C_{a_R} u_{b_R} \sim (1\,,\,1,-1)\times (3\,,\,1\,,\,2/3) \sim (3\,,\,1\,,\,-1/3),
    \label{LQinter}
    \end{eqnarray}
where in the singlet of  quarks above we are considering  the exotic ones. According to these bilinears the 331RHN supports the following set of leptoquarks:
\begin{eqnarray}
  &&  S_8 \sim (3\,,\, 8\,,\, 1/3),\,\,\,\,\,\,\,\, S_6 \sim (3\,,\,6\,,\,0),\nonumber \\ 
  && S_3 \sim(3\,,\ 3\,,\,0),\,\,\,\,\,\,\,\,\,\,\,\,\,\,\,\,\,\,\,\tilde S_3 \sim (3\,,3\,,\,1),\nonumber \\
  &&S \sim (3\,,\,1\,,\, 1/3),\,\,\,\,\,\,\,\,\,\,\tilde S \sim (3\,,\,1\,,\,4/3).
  \label{LQrep}
\end{eqnarray}

As we see, we have six categories of leptoquarks  with each one of them leading to a complex phenomenology in regard to flavor physics. 

According with the bilinears above, with the leptoquarks $S_8$ and $S_6$ we compose the following Yukawa interactions,
\begin{eqnarray}
    {\mathcal{L}}\supset  y^{LQ}_{a3} \bar L^C_{a_L} S_6 Q_{3_L} + y^{LQ}_{ai} \bar L^C_{a_L} S_8 Q_{i_L}  + H.c.,
\end{eqnarray}
where the $S_8$ and $S_6$ components are given by
\begin{eqnarray}
S_6=\left(\begin{array}{ccc}
\, \tilde R_1^{-5/3} & R_2^{-2/3} & R_3^{-5/3} \\
\newline \\
 R_2^{-2/3} &\,\tilde R_2^{-2/3}  & R_4^{-5/3} \\
\newline \\
 R_3^{-5/3}& R_4^{-5/3} &  \, \tilde R_3^{-1/3}  \end{array}\right)\,\,,\,\, S_8=\left(\begin{array}{ccc}
\, \tilde T_1^{-2/3} & T_2^{-5/3} & T_3^{-2/3} \\
\newline \\
 T_4^{-2/3} &\,\tilde T_2^{-5/3}  & T_4^{-2/3} \\
\newline \\
 T_5^{4/3}& T_6^{1/3} &  \, \tilde T_3^{4/3}  \end{array}\right).
\label{octeto} 
\end{eqnarray}

The triplet  $S_3$  and $\tilde S_3$ are  composed by the following set of leptoquarks,
\begin{eqnarray}
S_3 = \left (
\begin{array}{c}
S_3^{+1/3} \\
S_3^{+2/3} \\
S_3^{\prime +1/3}
\end{array}
\right )_L\,,
&&\tilde{S_3} = \left (
\begin{array}{c}
\tilde S_3^{+2/3} \\
\tilde S_3^{+4/3} \\
\tilde S_3^{\prime +2/3}
\end{array}
\right )_L
\label{tripletLQ} 
\end{eqnarray}
The third components of these triplets have $F(S^{\prime 1/3}_3 \,,\, \tilde S_3^{\prime 2/3})=-2$ while the other components have $F=0$ where $F=3B+L$. Their Yukawa interactions are given by
\begin{eqnarray}
    {\mathcal{L}}\supset \bar Y^{LQ}_{a3} \epsilon_{ijk} (\bar L^C_{a_L})_i (S_3)_j (Q_{3_L})_k +\tilde Y^{LQ}_{ab}\bar L_{a_L} \tilde S_3 d_{b_R}  + H.c.
\end{eqnarray}
Concerning the allowed singlets of leptoquarks, we  have that $\tilde S$  has $F=-2$ and, like the triplets above, do not leads to the flip of chirality. This means that they do not generate positive and robust contributions to the $g-2$ of the muon. 

The last leptoquark is the singlet $S$. It has $F=-2$ and compose the following Yukawa interactions with quarks and leptons,
\begin{equation}
  {\cal L} \supset \tilde{g}^{LQ}_{ia}\bar Q^C_{i_L} L_{a_L}S  + h^{LQ}_{ab}\bar u^C_{a_R} e_{b_R} S + H.c.,
  \label{LQS}
\end{equation}
with $i=1\,,\,2$ and $a,b=1,2,3$. Opening  the first term above we get,
\begin{equation}
 \bar Q^C_{i_L} L_{a_L}=\bar d^C_{i_L}  \nu_{a_L}-\bar u^C_{i_L} e_{a_L}+ \bar \nu_{a_R} d^{\prime}_{i_L}.  
\end{equation}
Observe that according with the Yukawa interactions above we identify that $S$  generates flip of chirality. It is due to this fact that from now on  we focus our attention exclusively in developing some physical aspects of this leptoquark.  

For reasons of simplification, and without any loss for the model,  we make the following set of assumptions: charged leptons, neutrinos, right-handed quarks and the new quarks come in a diagonal basis. With this  we obtain the Yukawa interactions among mass eigenstates of leptons and quarks with the leptoquark $S$,
\begin{equation}
  {\cal L}=\bar{\hat{u}}^C_{a}(g^{LQ}_{ab}P_L + h^{LQ}_{ab}P_R)e_b S + \bar{\hat{d}}^C_{a_L}g^{\prime LQ}_{ab} \nu_{a_L}S +  \bar \nu_{a_R}\tilde{g}^{LQ}_{ia}d^{\prime}_{i_L}S +  H.c.,
  \label{QS-Y}
\end{equation}
with $g^{LQ}_{ab}=-\tilde{g}^{LQ}_{ib}(V^u_L)_{bi}$ and $g^{\prime LQ}_{ab}=\tilde{g}^{LQ}_{ib}(V^d_L)_{bi}$. The physical quarks $\hat{u}_{a}=(u\,,\, c\,,\, t)$ and $\hat{d}_{a}=(d\,,\, s\,,\, b)$  are related to the symmetric  ones  by means of the transformations $\hat u_{L}=V^{\dagger u}_{L} u_{L}$ and $\hat d_{L}=V^{\dagger d}_{L} d_{L}$. The first term in Eq. (\ref{QS-Y}) is the one that generates the flip of chirality among the interaction of $S$ with the charged fermions. The interest in this interaction is because it provides a positive and  robust contribution to the $g-2$ of the muon as we see below.

As we sad in the introduction, the 331RHN with its original particle content, and even modified to implement the inverse seesaw mechanism, is not able to accommodate the current muon $g-2$. But now with the introduction of singlet of leptoquark $S$ we may cure this deficiency of the model. For this we made a brief review the contributions of the original particle content of the model to the muon $g-2$ at one-loop.

\section{review of the one loop contribution to $\Delta a_\mu $ in the 331RHN and the contribution of the leptoquark S}
\subsection{331RHN original particle content contribution to $\Delta a_\mu $}
We review the main contributions at one loop to the muon anomalous magnetic moment $a_\mu =\frac{(g-2)_\mu}{2}$ in the 331RHN. For this we follow Ref. \cite{deJesus:2020ngn} where it is took $v_\eta=v_\rho=v=174$ GeV. We are interested  in the discrepancy  $\Delta a_\mu$ produced by the new charged gauge boson $W^{\prime \pm}$, $Z^{\prime}$, the single charged scalar $h^+_{1}$ , the neutral scalars $h_2$  and the pseudo Goldstone $A$\footnote{The charged scalar $h^+_2$ and the neutral scalar $H$ do not contribute to $(g-2)_\mu$\cite{Pinheiro:2022bcs}}.

The  interactions of interest  are given by
\begin{equation}
\label{doublyminimal}
{\cal L}^{CC}_l= - \frac{g}{2\sqrt{2}}\left[
\overline{\nu^c_R}\, \gamma^\mu (1- \gamma_5) \bar{l}\, W^{\prime -}_\mu  \right],
\end{equation}
\begin{equation}{\cal L}^{NC} =
\bar{f}\, \gamma^{\mu} [g_{V}(f) + g_{A}(f)\gamma_5]\, f\,
Z'_{\mu}, \label{ncm}
\end{equation}
\begin{equation}
    {\cal L}^{h^+_1}_Y= \sqrt{2}\frac{m_l}{v}\bar \nu_{l_L} \l_R h^+_1 ,
\end{equation}
\begin{equation}
{\cal L}^{h_2+A} =\frac{ m_{\mu}}{ \sqrt{2}v}\bar{\mu}\, \mu h_2 + i \frac{m_\mu}{v} \bar \mu \gamma_5 \mu A,
\label{neutralSca}
\end{equation}
where $g_{V}$ and $g_{A}$ are the vector and axial coupling constants and are given by,
\begin{equation}
g'_{V}(l) = \frac{g}{4 c_W} \frac{(1 -
4 s_W^2)}{\sqrt{3-4s_W^2}},\
g'_{A}(l) = -\frac{g}{4 c_W \sqrt{3-4s_W^2}},
\label{hsr}
\end{equation}
The Feymnan diagrams displaying the contributions of these interactions to $(g-2)_\mu $  are showed in FIG. 1.
\begin{figure}
    \centering
\includegraphics[scale=0.5]{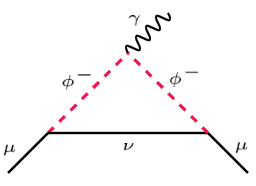}
\includegraphics[scale=0.5]{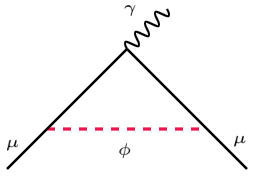}
\includegraphics[scale=0.5]{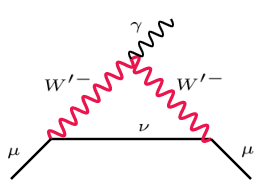}
\includegraphics[scale=0.5]{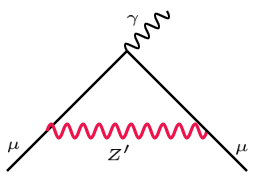}
    \caption{One-loop contributions to the $g-2$ of the muon in the 331RHN where $\phi\,\,\mbox{and}\,\, \phi^-$  represent the neutral and charged scalar contributions}
    \label{fig:pseudoscalarV1_2}
\end{figure}

The  evaluation of these contributions at one-loop to $\Delta a_\mu$ generated by these set of particles follow the procedure of Ref. \cite{deJesus:2020ngn} . To avoid being repetitive we simply quote them in the heavy mediator regime. With that being said, we have  the following contributions of the 331RHN to the  $\Delta a_\mu$,
  
  \begin{eqnarray}
      \Delta a_\mu(\nu,W^\prime) \simeq \frac{1}{4\pi^2}\frac{m_\mu^2}{M_{W^\prime}^2} \left|\frac{g}{2\sqrt{2}}\right|^2   \left( \frac{5}{3}  \right),
 \label{arhn1}
  \end{eqnarray}
  
\begin{eqnarray} 
	\Delta a_\mu\left(\mu,Z^\prime\right) \simeq \frac{-1}{4\pi^2}\frac{m_\mu^2}{M_{Z^\prime}^2} \frac{1}{3}\left|-\frac{g}{4 c_W \sqrt{3-4s_W^2}}\right|^2   \left[ -\left|  1 -
4 s_W^2\right|^2  +5  \right],
\end{eqnarray}
    
\begin{eqnarray}
	\Delta a_\mu(h^+_1) \simeq  \frac{-1}{4\pi^2}\frac{m_\mu^2}{M_{h^+_1}^2}  \left|\frac{m_\mu\sqrt{2}}{2 v_\eta}\right|^2  \frac{1}{6},
\end{eqnarray}
    
\begin{eqnarray} 
\Delta a_\mu(h_2) \simeq \frac{1}{4\pi^2}\frac{m_\mu^2}{M_{h_2}^2}  \left(\frac{m_\mu\sqrt{2}}{2 v_\eta}\right)^2 \left[ \frac{1}{6} + \left(\frac{3}{4} + \log\left(\frac{m_\mu}{M_{h_2}}\right)\right)\right].
\end{eqnarray}

\begin{eqnarray}
    \Delta a_\mu(A)= - \frac{m_\mu^2}{8\pi^2M_A^2}\Bigg(\frac{g A_\mu m_\mu}{2 M_W}\Bigg)^2 H\Bigg(\frac{m_\mu^2}{M_A^2} \Bigg),
 \label{arhn5}
\end{eqnarray}
\begin{figure}[t]
\centering
\includegraphics[width=0.5\columnwidth]{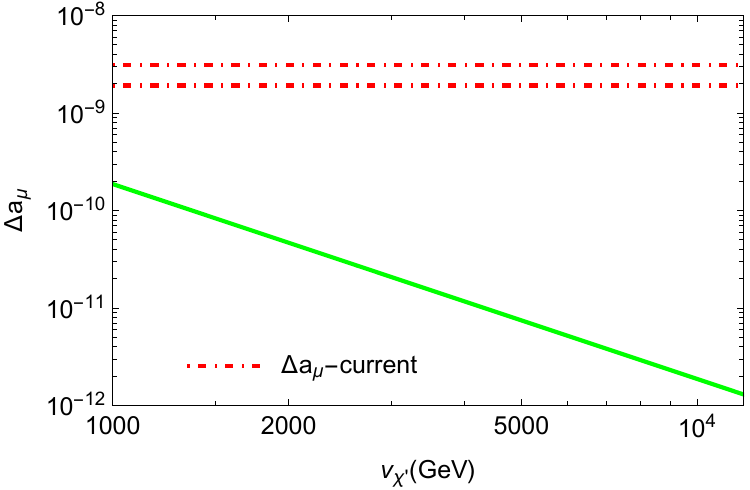}
\caption{Overall contribution to $\Delta a_\mu$ from Eqs.(\ref{arhn1})-(\ref{arhn5}) to the 331RHN  model. The dash-dotted bands in red are delimited by the current $1\sigma$ bound. }
\label{a331rhn}
\end{figure}
with  $H(y)=\int_0^1 \frac{dx x^3}{1-x+x^2y}$ where $y=\frac{m^2_\mu}{M^2_A}$ . In summary, the 331RHN yields five contributions to $\Delta a_\mu$ with only $W^{\prime}$ and $h_2$ contributions  being positive. In view of this, according to Ref. \cite{DeJesus:2020yqx}, the only way the one-loop contribution would accommodate the current prediction of the $\Delta a_\mu$ was if the positive contributions of the $W^{\prime}$  and $h_2$ could surpass the other  ones. This could happen for $W^{\prime}$ having mass at the electroweak scale which is completely discarded since current LHC bounds demands the gauge bosons of the 331RHN have mass at the TeV scale. To be precise LHC demands $m_{Z^{\prime}}$ with mass in the range $(3-4)$ TeVs\cite{Coutinho:2013lta}, depending if $Z^{\prime}$ may decay in the new quarks or don´t\cite{Alves:2022hcp}. This implies $v_{\chi^{\prime}}$ in the range $10-12$TeVs.  In FIG.\ref{a331rhn}   we present the prediction of the 331RHN to $\Delta a_\mu$ in function of  $v_\chi^{\prime}$ varying in the range $1-12$ TeVs. As we can see in FIG. \ref{a331rhn}   one-loop contributions to $\Delta a_\mu$ in the 331RHN are not able to accommodate the current $\Delta a_\mu$. Observe that for $v_{\chi^{\prime}}$ around $10^4$ GeV the prediction to $\Delta a _\mu$ lies three order of magnitude below the experimental prediction. Thus the one loop contribution to $\Delta a_\mu$ in the 331RHN is practically neglegible.  As sad in the introduction, it is hard to find  modifications of the 331RHN that accommodate the current $\Delta a_\mu$ value. A simple route to accommodate  $\Delta a_\mu$ into the 331RHN is via the introduction of vectorlike leptons and inert scalar triplets to the model\cite{DeJesus:2020yqx}.  What we are going to show next is that the addition of only one singlet of leptoquark resolves the $(g-2)_\mu$ anomaly.

\subsection{The contribution of $S$ to $\Delta a_\mu$ in the 331RHN}
\begin{figure}[t]
\centering
\includegraphics[width=0.6\columnwidth]{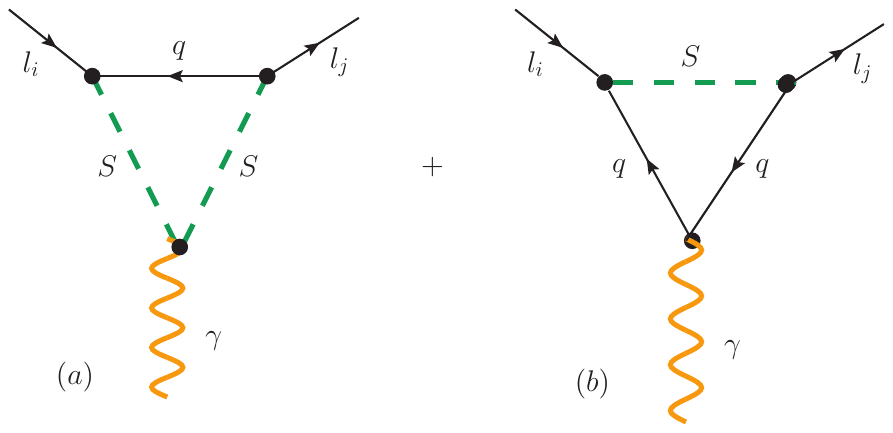}
\vspace*{-9cm}
\caption{One-loop diagrams contributing to $\Delta a_\mu$ and $\mu \to e \gamma$  decay  induced by S leptoquarks: To the contributions to $\Delta a_\mu$ in this figure $l_i=l_j = l_2 =\mu$, while to muon decay  we have  $l_i=l_2$ and  $l_j=l_1=e$.}
\label{Scont}
\end{figure}
The one-loop contributions to $\Delta a_\mu$ due to the leptoquark $S$, depicted in the FIG. \ref{Scont} and in  according to the interactions in Eq. (\ref{QS-Y}) are given by\cite{Dorsner:2016wpm,Athron:2021iuf}
\begin{equation}
 \Delta a_\mu^{LQ}=\frac{m^2_\mu}{48 \pi^2 m^2_S}f(m_\mu)\left( \frac{m_q}{m_\mu}g^{LQ}_{a2}h^{LQ}_{a2}L_1(x_q)+\frac{(g^{LQ}_{a2})^2 +(h^{LQ}_{a2})^2}{4}L_2(x_q) \right), 
 \label{LQflip}
\end{equation}
where $x_q=\frac{m^2_q}{m^2_{S}}$ and 
\begin{eqnarray}
&& L_1(x)=4F_F(x)-F_C(x)\,,\,\,\,\,\, L_2(x)=2F_E(x)-F_B(x) \nonumber \\
&& F_F(x)=\frac{3(-3+4x-x^2-2lnx)}{2(1-x)^3}\,,\,\,\,\,\, F_E(x)=\frac{2(2+3x-6x^2+x^3+6xlnx)}{(1-x)^4}, \nonumber \\
&& F_C(x)=\frac{3(1-x^2+2xlnx)}{(1-x)^3}\,,\,\,\,\,\,F_B(x)=\frac{2(1-6x+3x^2+2x^3-6x^2 lnx)}{(1-x)^4}. \nonumber \\
&& f(m_\mu) = ( 1 + \frac{4\alpha}{\pi}\ln(\frac{m_\mu}{m_{S}})).
\end{eqnarray}

\par We consider only the maximum contribution given by Eq. (\ref{LQflip}) which is due to the top quark ($a=3$ in Eq. (\ref{LQflip})). The free parameters involved in such contributions are the Yukawa couplings $g^{LQ}_{32}$, $h^{LQ}_{32}$ and the mass of $S$ which we called $m_S$.  The current LHC bounds  demands the lowest  leptoquark to have mass in the range $O(0.98–1.73)TeV$, depending on the spin of  leptoquarks  and its couplings to SM fermions\cite{CMS:2024bnj}. Here we adopt the current assumption $m_{S} = 1.8TeV$ and extract constraint on the Yukawa couplings $g^{LQ}_{32}$ and $h^{LQ}_{32}$.

\par
\begin{figure}[t]
\centering
\includegraphics[width=0.5\columnwidth]{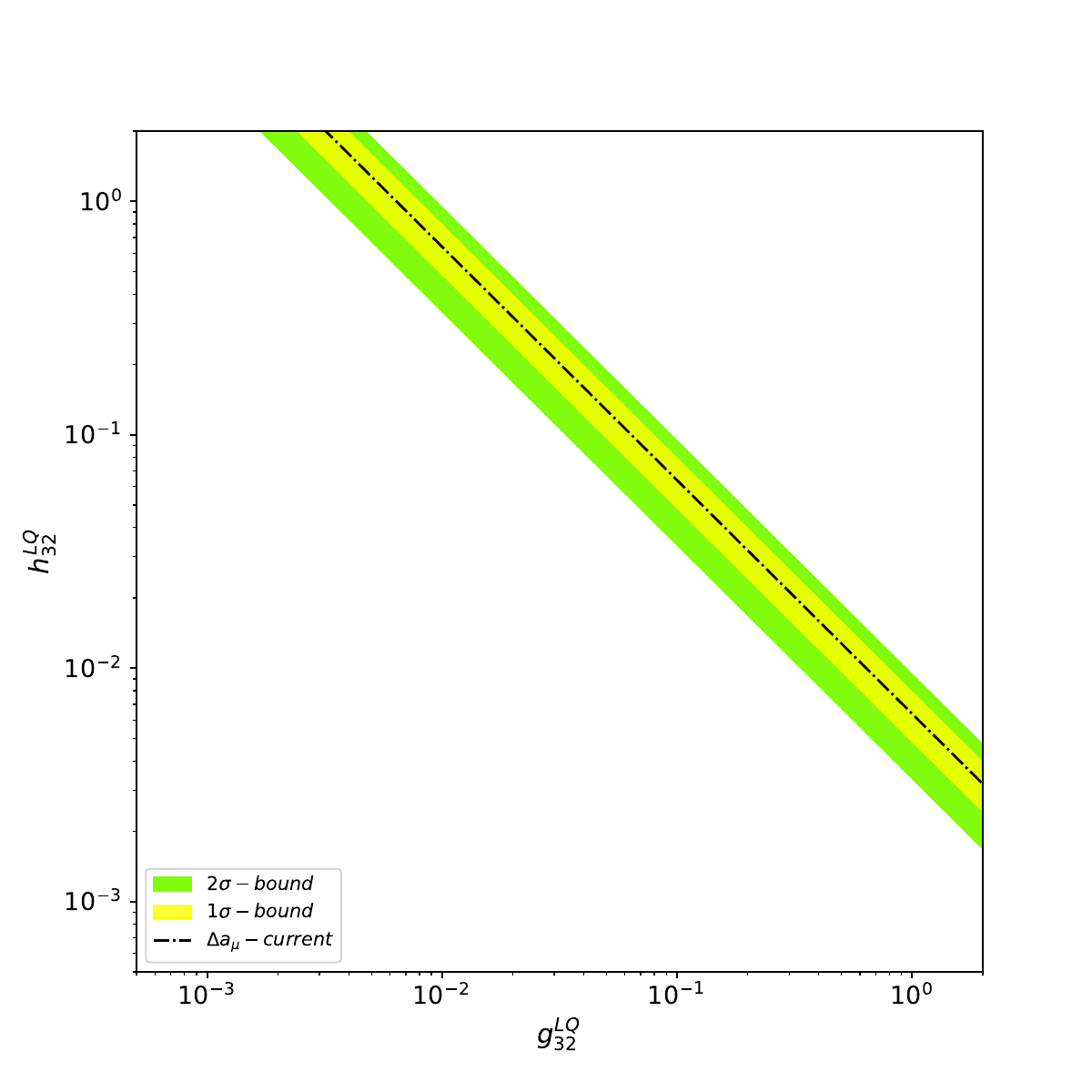}
\caption{ In this figure we show in parameter space ($g^{LQ}_{32}-h^{LQ}_{32}$)  the  allowed  values to $\Delta a_{\mu}$. The dash-dotted line corresponds to central value obtained to $\Delta a_{\mu}$. The respective bands $1(2)\sigma$ are represented in yellow(green)}
\label{delta-S}
\end{figure}

 Our numerical result is displayed in  FIG. \ref{delta-S}. As we can see there  the existence   of a simple singlet of leptoquark $S$  presenting flip of chirality  give positive and robust contributions to $\Delta a_\mu $  explaining easily the observed value of  $\Delta a_\mu$. That is a very nice result in what concern the development of the 331RHN. In the figure  the dash-dotted line corresponds to central value  $\Delta a_\mu = 25.1\times 10^{-10}$. The Yellow(Green) bands  corresponds to the current $1\sigma(2\sigma)$ bounds found by requiring  $\Delta a_\mu \lesssim 59 \times 10^{-11}$. In the next section, we complement this result with the study of the decay $\mu \rightarrow e\gamma$ once the contributions are the same being necessary just to take of-shell photon in FIG. \ref{Scont}.

\section{Contributions of $S$  leptoquarks to $\mu \to e \gamma$ decay}

\par  In the FIG. \ref{Scont}  we show the contributions  of $S$  leptoquarks   to the muon decay  $\mu \to e \gamma$. These contributions are 
similar to the  corrections  of $S$ to $\Delta a_\mu$, and in this Feynman diagrams we consider only the  chirality enhanced
terms, given by Eq.(\ref{LQS}), that explain  the  observed $\Delta a_\mu$. To this decay the branching ratio presents the following form\cite{Br:1996hm}
\begin{equation}
BR(\mu \to e \gamma) = \frac{m^5_{\mu}}{16\pi\Gamma_\mu}\left(|A^L_2|^2 + |A^R_2|^2\right)
\label{BRmu}
\end{equation}
\noindent  where $\Gamma_\mu = 2.996\times 10^{-19}GeV$  and the structure of the dipole form factors $A^2_{L,R}$ are given by 
\begin{eqnarray}
&& A^L_2 = -\frac{1}{16\pi^2}\frac{e}{6m^2_S}\left(\frac{m_q}{m_\mu}g^{LQ}_{a1}h^{LQ}_{a2}L_1(x_q) + \frac{1}{4}h^{LQ}_{a1}h^{LQ}_{a2}L_2(x_q)\right), \nonumber \\
&& A^R_2 = -\frac{1}{16\pi^2}\frac{e}{6m^2_S}\left(\frac{m_q}{m_\mu}h^{LQ}_{a1}g^{LQ}_{a2}L_1(x_q) + \frac{1}{4}g^{LQ}_{a1}g^{LQ}_{a2}L_2(x_q)\right).
\label{FFA2}
\end{eqnarray}

\noindent Following the same line of the previous section, we will consider  the maximum contribution to $\mu \to e \gamma$ decay  that is due to the top quark ($a=3$).

\par Considering the comparison with the experimental results to $\Delta a_\mu$, in the FIG. \ref{delta-S} we show  the bounds on products of 
$g^{LQ}_{32}$ and $h^{LQ}_{32}$ couplings in parameter space. In this figure the band in  Green  restricts the product of  these couplings 
, i. e. the left and handed  leptoquarks couplings of top with the muon. As consequence the $2\sigma$ band  results in the following allowed range of possible 
couplings
\begin{equation}
 3.1\times 10^{-3} <  g^{LQ}_{32}*h^{LQ}_{32} < 9.3\times 10^{-3}. 
\label{Lcoupl}
\end{equation}

\par In the Eq.(\ref{FFA2}) the left and handed  leptoquarks couplings of top with the electron are $g^{LQ}_{31}$ and $h^{LQ}_{31}$. In order to find
the allowed parameter space in the $(g^{LQ}_{31}$  -  $h^{LQ}_{31})$ plane,  currently permitted by experimental bounds\cite{MEGI}\cite{MEGII}, we can consider 
the above result, combined with the expression to the branching ratio given by Eq.(\ref{BRmu}). 

\par  Therefore,  considering the  Eqs.(\ref{BRmu}), (\ref{FFA2})  and (\ref{Lcoupl}) we can write the following inequality

\begin{figure}[t]
\centering
\includegraphics[width=0.8\columnwidth]{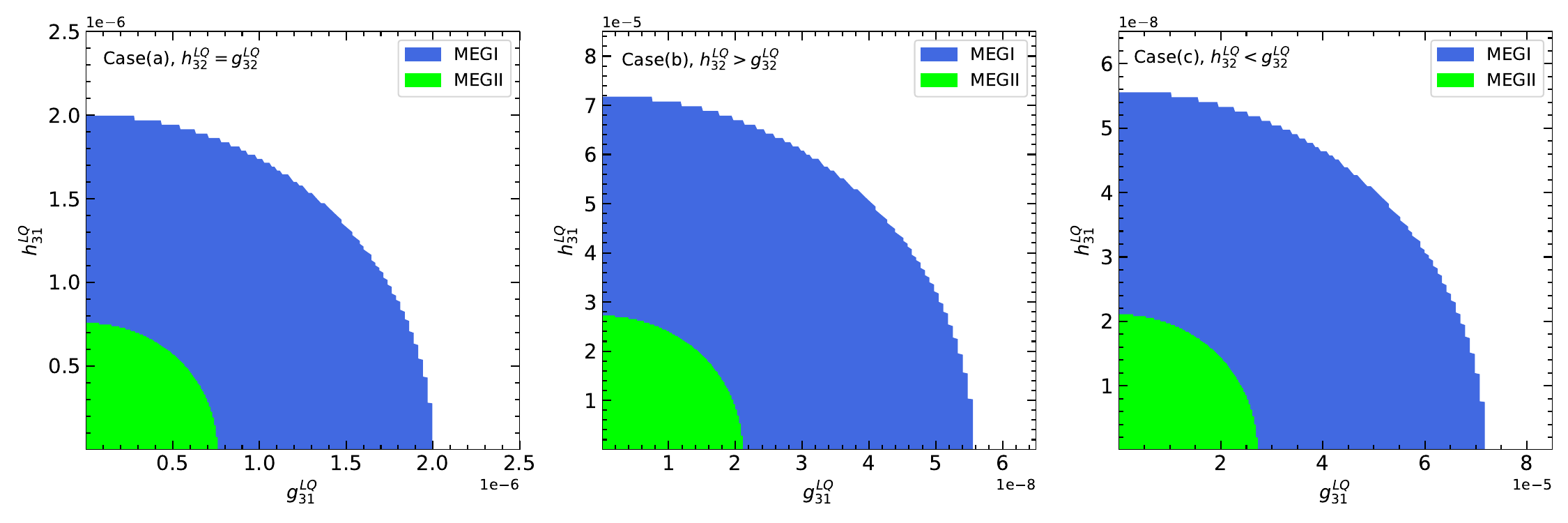}
\caption{Allowed parameter regions for the $\mu \to e \gamma$ decay, assuming that $\Delta a_\mu$ is explained by  top quark chirally enhanced
term. The contextualization of the curves behavior are described in the text.}
\label{decay}
\end{figure} 

 \begin{equation}
 \left(|h^{LQ}_{32}|^2|h^{LQ}_{31}|^2 + |g^{LQ}_{32}|^2 |g^{LQ}_{31}|^2 \right)  < 3\times10^{-2}BR(\mu \to e \gamma).
\label{31LQ}
\end{equation}
 \noindent  In order to generate the parameter region allowed, we will consider the following possibilities of top couplings
 \begin{eqnarray}
&& {\rm (a) Symmetric\,\,leptoquark\,\,coupling:}\,\, h^{LQ}_{32} = g^{LQ}_{32},\nonumber \\
&& {\rm (b) Asymmetric\,\,leptoquark\,\, coupling:}\,\,h^{LQ}_{32} > g^{LQ}_{32},\nonumber \\
&& {\rm (c) Asymmetric\,\,leptoquark\,\, coupling:}\,\,h^{LQ}_{32} < g^{LQ}_{32}. 
\label{FFA20}
\end{eqnarray}
 \par Assuming the  lower bound $(3.1\times 10^{-3} <  g^{LQ}_{32}*h^{LQ}_{32})$, we obtain to Case (a) ($|h^{LQ}_{32}|^2 = |g^{LQ}_{32}|^2 > 3.1\times 10^{-3}$), while that to $h^{LQ}_{32}=2$, we have to Case (b) $(g^{LQ}_{32} > 1.55 \times 10^{-3})$ and similarly to $g^{LQ}_{32}=2$  the  Case (c) corresponds to  $(h^{LQ}_{32} > 1.55 \times 10^{-3})$. 
 
 \par Currently the experimental limits  to $BR(\mu \to e \gamma)$ correspond to $BR(\mu \to e \gamma)=4.2 \times 10^{-13}$\cite{MEGI} and $BR(\mu \to e \gamma)=6\times 10^{-14}$\cite{MEGII}. In FIG. \ref{decay} we show in  the plane $(g^{LQ}_{31}$  -  $h^{LQ}_{31})$  the  allowed  values to $g^{LQ}_{31}*h^{LQ}_{31}$, assuming  the considerations indicated in the paragraph above.

 \par In FIG. (\ref{decay}a), corresponding to the symmetric case,  the region filled in blue leads to the upper bond to the  allowed  values of  couplings as 
 $g^{LQ}_{31}*h^{LQ}_{31} < 2.056 \times 10^{-12}$,  while for region filled in green  we have an more restrictive  upper bond $g^{LQ}_{31}*h^{LQ}_{31} < 2.93 \times 10^{-13}$. 
 \par For the asymmetric cases corresponding to FIG. (\ref{decay}b) and (\ref{decay}c), the limits obtained are smaller compared to case (a), so that they are within the region covered in this situation. Considering the combination of results presented in these figures, that illustrate the  allowed  values to
S leptoquarks  couplings  permitted by actual experimental bounds, we found that leptoquarks couplings to electrons must be more than 4 orders of magnitude smaller than the corresponding couplings to muons.


 \section{discussion}
Anomalies cancellation inside 3-3-1 models demand that at least one family of quarks transforms differently form the other two\cite{Frampton:1992wt,Ng:1992st,Liu:1993gy}. This means that for each version of 3-3-1 models  three are three variants as discussed in Ref.\cite{Oliveira:2022dav,Oliveira:2022vjo} . Each variant has its proper implications in flavor physics. However, in adding leptoquarks to the 331RHN, besides the Yukawa interactions among leptoquarks and fermions  seems to discriminate families, as expressed in Eq. (\ref{LQS}), when we go the to mass eigenstates we see that leptoquarks do not distinguish families, as expressed in Eq. (\ref{QS-Y}). This is so because multiplet of leptoquarks  do not contribute to the masses of the fermions after spontaneous breaking of the symmetry. Then we can always redefine the Yukawa couplings by means of the mixing  matrix  that transform fermions  eigenstate of symmetry into the mass eigenstates. Thus the fact we have chosen the third family transforming as triplet and the first and second ones transforming as anti-triplet was a mere question of convenience.  

Another point we would like to call the attention is that while the standard model supports  two representaion of leptoquarks inducing flip of chirality\cite{ColuccioLeskow:2016dox}, namely a singlet and a doublet of leptoquarks, in the 331RHN, besides it supports a more extensive set of multiplet of leptoquarks (octet, sextet, triplet and singlet), however only a specific leptoquark transforming as singlet by the 3-3-1 symmetry leads to the flip of chirality. Nevertheless, such singlet  is sufficient to do the job of explaining $\Delta a _\mu$.

Observe that the leptoquark $S$ in eq. (\ref{QS-Y}) enhances the contribution to $(g-2)_\mu$ by a factor $\frac{m_t}{m_\mu}$. However if we switch on the VEVs of $\chi^0$ and $\eta^{\prime 0}$ we have, as main consequence, a mixing among the standard quarks with the new ones. As main implication to $(g-2)_\mu$ we will have an enhancement of $\Delta a_\mu$ by a factor of $\frac{m_{u^{\prime}}}{m_\mu}$. A $m_{u^{\prime}}$ at TeV scale would give a very robust contribution to $\Delta a _\mu$ rendering constraints in the mixing among standard and new quarks which translate in constraints on the VEVs $v_{\chi}$ and $v_{\eta^{\prime}}$. 

 In FIG. \ref{Scont} we took $q=t$ which implies an enhancement of the contributions to  $\Delta a_\mu$ by a factor of $\frac{m_t}{m_\mu}$ . This is the reason why people consider in the standard model only the interaction of leptoquarks with the third generation of quarks. We can not do this here because the symmetries do not allow. The leptoquark $S$ has an intriguing set of Yukawa interactions as we can see in Eq. (\ref{LQS}). The interaction  with the quark top arises due to the mixing among the quarks given by the mixing matrix $V^{u,d}_L$. 

 \section{conclusion}
 In this work we introduced leptoquarks into the 3-3-1 model with right-handed neutrinos with the aim of resolving the problem of the $\Delta a_\mu$ anomaly since the original version of the model gives a neglegigle contribution to $\Delta a_\mu$. We saw that the model supports multiplet of leptoquarks in the representation of octet, sextet, triplet and singlet. We determined the particle content of these multiplets and presented they Yukawa interactions with the fermions of the model.

 Besides this large spectrum of leptoquarks, only a specific singlet $S$ is capable of generate a robust contribution to $\Delta a_\mu$ since only it leads to flip of chirality among the fermions. We then developed the main aspects of the leptoquark $S$ in what concern its contributions to $\Delta a_\mu$ and to the muon decay channel $\mu \rightarrow e \gamma$. Our main result is that singlet of leptoquark is not only  viable but also the most economical extension of the 3-3-1 model that explain the $\Delta a_\mu$ anomaly.

\section*{Acknowledgments}
C.A.S.P  was supported by the CNPq research grants No. 311936/2021-0.
\bibliography{bibliography}

\end{document}